\documentclass[
 reprint,
superscriptaddress,
 amsmath,amssymb,
 aps,
]{revtex4-2}

\usepackage[margin=2.5cm]{geometry}
\usepackage{amsmath,amssymb}
\usepackage{graphicx}
\usepackage[colorlinks=true,linkcolor=blue,citecolor=blue,urlcolor=blue]{hyperref}
\usepackage{booktabs}
\usepackage{physics}
\usepackage{orcidlink}

\newcommand{\Gc}{G_{\mathrm{c}}}
\newcommand{\Geff}{G_{\mathrm{eff}}}
\newcommand{\Dcal}{D_{\mathrm{cal}}}
\newcommand{\Dtrue}{D_{\mathrm{true}}}

\begin{document}

\preprint{APS/123-QED}

%\title{The self-oscillation threshold of a digitally frequency-locked NV-centre magnetometer as a probe of transduction changes:\\ a testbed study for sensing with autonomous devices}% Force line breaks with \\

\title{Critical Sensing with Autonomous Devices: \\The Self-Oscillation Threshold of a Frequency-Locked
NV-Centre Magnetometer}% Force line breaks with \\

\author{Joan Toledo Aguilera\,\orcidlink{0009-0004-0914-2053}}%
\email{joan.toledo@upct.es}
\affiliation{Universidad Politécnica de Cartagena member of European University of Technology EUT+, Research Group of Quantum Technologies, Cartagena E-30202, Spain}

\author{Gonzalo Reina Rivero\,\orcidlink{0000-0003-4219-2306}}%
\email{gonzalo.reina@upct.es}
\affiliation{Universidad Politécnica de Cartagena member of European University of Technology EUT+, Research Group of Quantum Technologies, Cartagena E-30202, Spain}

\author{Marcel Morillas-Rozas\,\orcidlink{0009-0005-4570-1016}}%
 \email{marcel.morillas@upct.es}
\affiliation{Universidad Politécnica de Cartagena member of European University of Technology EUT+, Research Group of Quantum Technologies, Cartagena E-30202, Spain}

\author{Javier Cerrillo\ \orcidlink{0000-0001-8372-9953}}%
 \email{javier.cerrillo@upct.es}
\affiliation{Universidad Politécnica de Cartagena member of European University of Technology EUT+, Research Group of Quantum Technologies, Cartagena E-30202, Spain}

\begin{abstract}
Feedback locking of a probe frequency to a spin resonance is the standard
operating mode of precision quantum sensors. Here we deliberately operate such a
lock outside its stable regime: a continuous-wave nitrogen-vacancy (NV)
ensemble magnetometer, frequency-modulation (FM) locked to one flank of its
optically detected magnetic resonance (ODMR), is driven through the flip
(period-doubling) bifurcation of its discrete feedback map by raising the
software loop gain $G$. Beyond a critical gain $\Gc$ the lock becomes a
self-sustained oscillator whose limit cycle is generated by the loop itself. We
derive the threshold condition $\Gc = 2\,\Dcal/\Dtrue$, which identifies the
measurable content of the threshold: the ratio of the transduction
slope of the ODMR lock-in signal at calibration time $D_{cal}$ to its value at present $D_{true}$. We present an identifiability analysis showing which physical parameters this single
scalar can and cannot distinguish, characterize the estimators of $\Gc$ under
realistic noise, and report measurements on our current setup: an
experimental bifurcation diagram with onset at $\Gc \approx 2$ as predicted for
a self-calibrated loop, sub-threshold critical fluctuations following the
predicted $\sqrt{G/(2-G)}$ divergence. 
\end{abstract}

\maketitle

\section{Introduction}
\label{sec:intro}

Quantum sensors based on the nitrogen-vacancy (NV) centres in diamond
translate magnetic fields into shifts of a spin-resonance frequency, read out
optically \cite{Degen2017,Barry2020}. In practice, precision measurements
can be performed via closed-loop operation: the probe frequency is continuously adjusted to
the resonance, and the error signal is the measurement. The design goal
of such loops in many quantum sensing applications is to improve aspects like dynamic range, avoiding saturation of the sensor signal when encountering large perturbations.

A distinct line of research uses the opposite regime: near a bifurcation, the
qualitative state of a driven nonlinear system responds sharply to small
parameter changes. This is the operating principle of the Josephson
bifurcation amplifier \cite{Siddiqi2004}, of a family of MEMS
bifurcation-based sensors \cite{MEMSunlock2023,MEMSmass2020}, and, in a
quantum-critical form, of parametric sensors operated near driven-dissipative
phase transitions \cite{DiCandia2023,PRXQ2025}. In these systems the
nonlinearity is \emph{physical} and acts before the dominant readout noise.

\begin{figure}[tb]
\centering
\includegraphics[width=0.95\columnwidth]{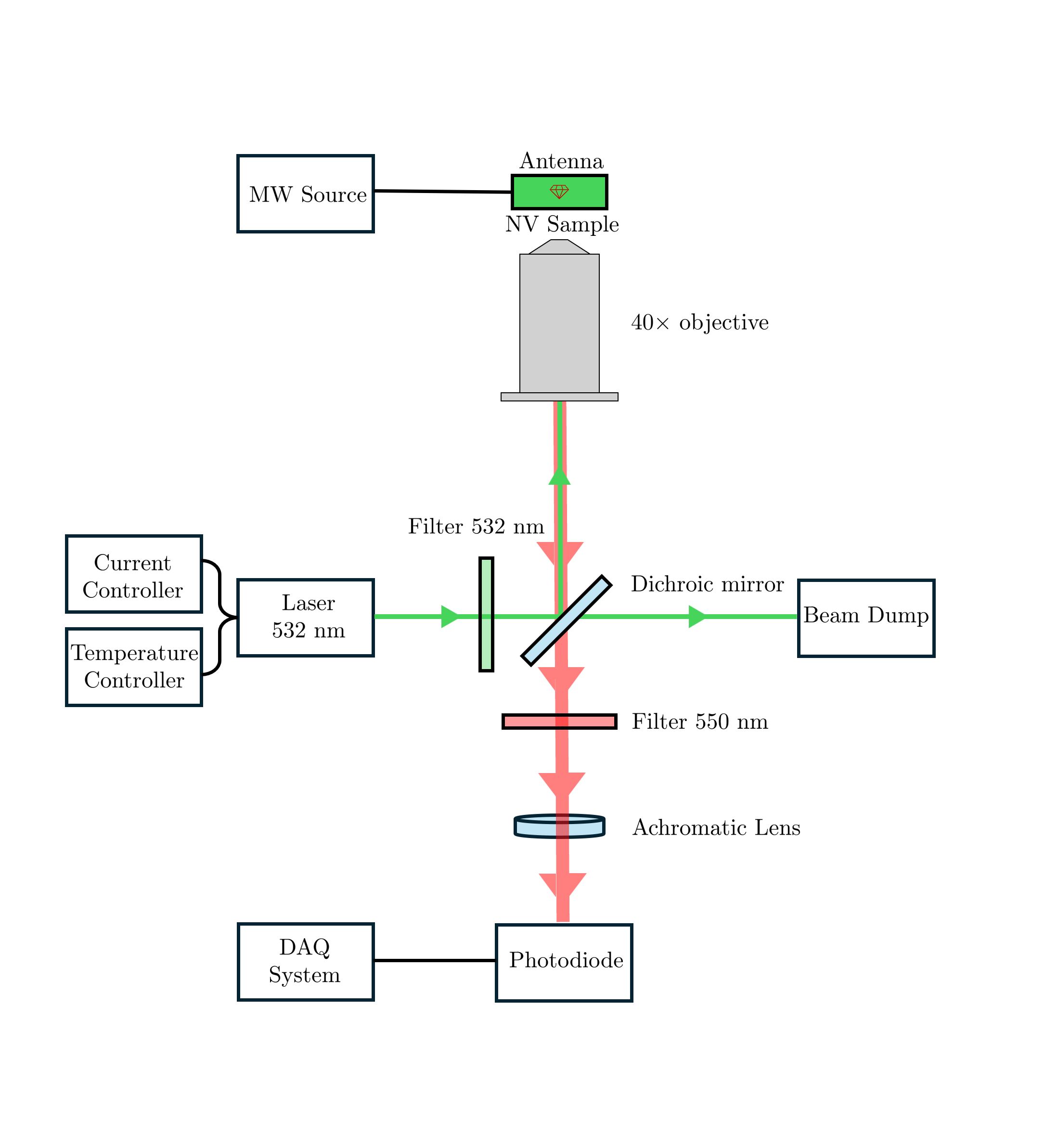}
\caption{Schematic of the continuous-wave ODMR magnetometer used in this
work.}
\label{fig:setup_scheme}
\end{figure}

The present work sits deliberately between these two traditions, where we study the idea of autonomous self-oscillating behaviour for potential quantum sensing applications.
We take the ordinary digital frequency lock of a CW NV magnetometer shown schematically in Fig.~\ref{fig:setup_scheme} and drive it through its own instability -- the flip (period-doubling) bifurcation familiar from sampled-data control systems \cite{Banerjee2001,Strogatz} -- turning the lock into an autonomous self-oscillator. We then study what kind of experimental parameters might affect the threshold gain $\Gc$. Our aims are
methodological: (i) to verify the quantitative dynamics of the loop
(threshold location, critical fluctuations, limit-cycle statistics) on real
hardware; (ii) to establish, by an explicit identifiability analysis, the
information content and the limits of $\Gc$ as an observable; and (iii) to
exercise, on an accessible classical platform, the estimators and statistical
tools (threshold extraction, two-time statistics) that physically autonomous
sensors -- where the nonlinearity precedes the readout noise -- will require.
We state at the outset, and quantify in Sec.~\ref{sec:discussion}, that
because the nonlinearity here is implemented in software \emph{after}
digitization, no metrological advantage over direct linear readout is possible
in this system; the value of the exercise is as a controlled testbed.

\section{Theory}
\label{sec:theory}

\begin{figure*}[tb]
\centering
\includegraphics[width=\textwidth]{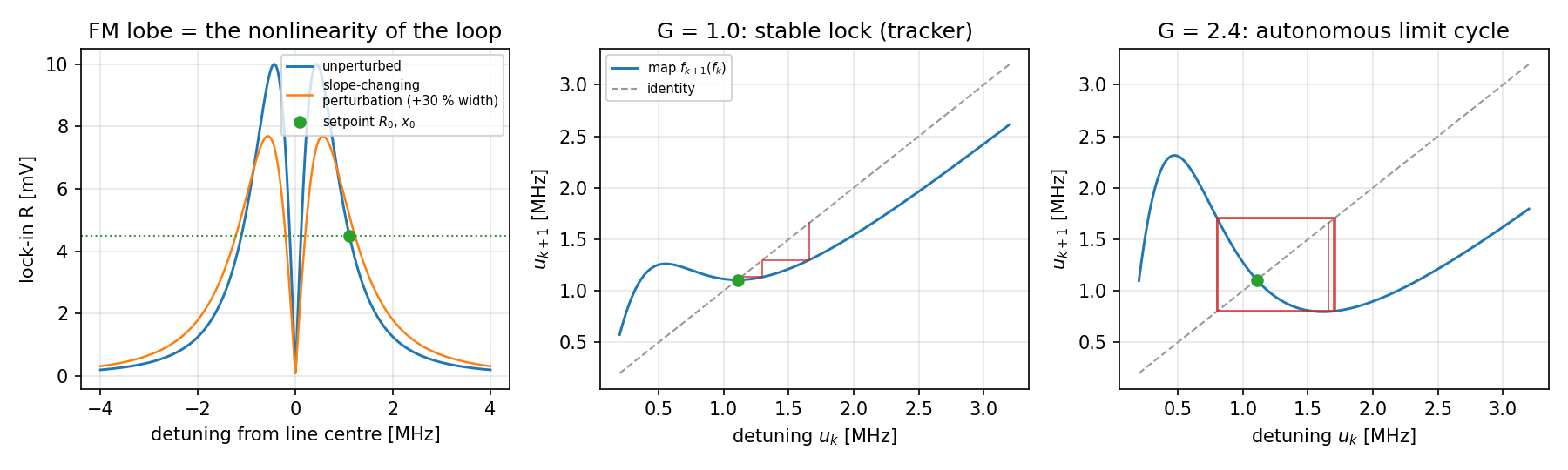}
\caption{The lock as a dynamical system (simulation with rig-like
parameters). Left: the FM lock-in lobe $R(f)$ is the nonlinearity of the map;
the working point sits on a flank. Middle and right: cobweb diagrams of
Eq.~\eqref{eq:map} below threshold (stable lock, $G=1$) and above threshold
($G=2.4$), where the orbit settles on a period-2 limit cycle.}
\label{fig:system}
\end{figure*}

\subsection{The frequency lock as a discrete map}
\label{sec:map}

The magnetometer is a CW ODMR setup in which the microwave (MW) frequency is modulated at $f_m = 5$~kHz and the photodiode signal is demodulated in software (magnitude lock-in, no phase reference), yielding a signal $R(f)$
proportional to the absolute derivative of the ODMR line: two lobes with a
null at the line centre. A working point $f^{*}$ is chosen on a lobe flank,
with setpoint $R_0 = R(f^{*})$ and calibrated slope
$\Dcal = (\mathrm{d}R/\mathrm{d}f)|_{f^{*}}$.

Once per lock cycle the loop measures $R$ at the current frequency $f_k$ and
applies a proportional correction,
\begin{equation}
f_{k+1} \;=\; f_k \;-\; \frac{G}{\Dcal}\,\bigl[R(f_k) - R_0\bigr],
\label{eq:map}
\end{equation}
where the dimensionless loop gain $G$ is a software parameter (the fraction of
the estimated frequency error corrected per cycle). Equation~\eqref{eq:map} is
a one-dimensional discrete map whose nonlinearity is the measured lobe shape
$R(f)$ itself (Fig.~\ref{fig:system}).

\begin{figure*}[tb]
\centering
\includegraphics[width=0.8\textwidth]{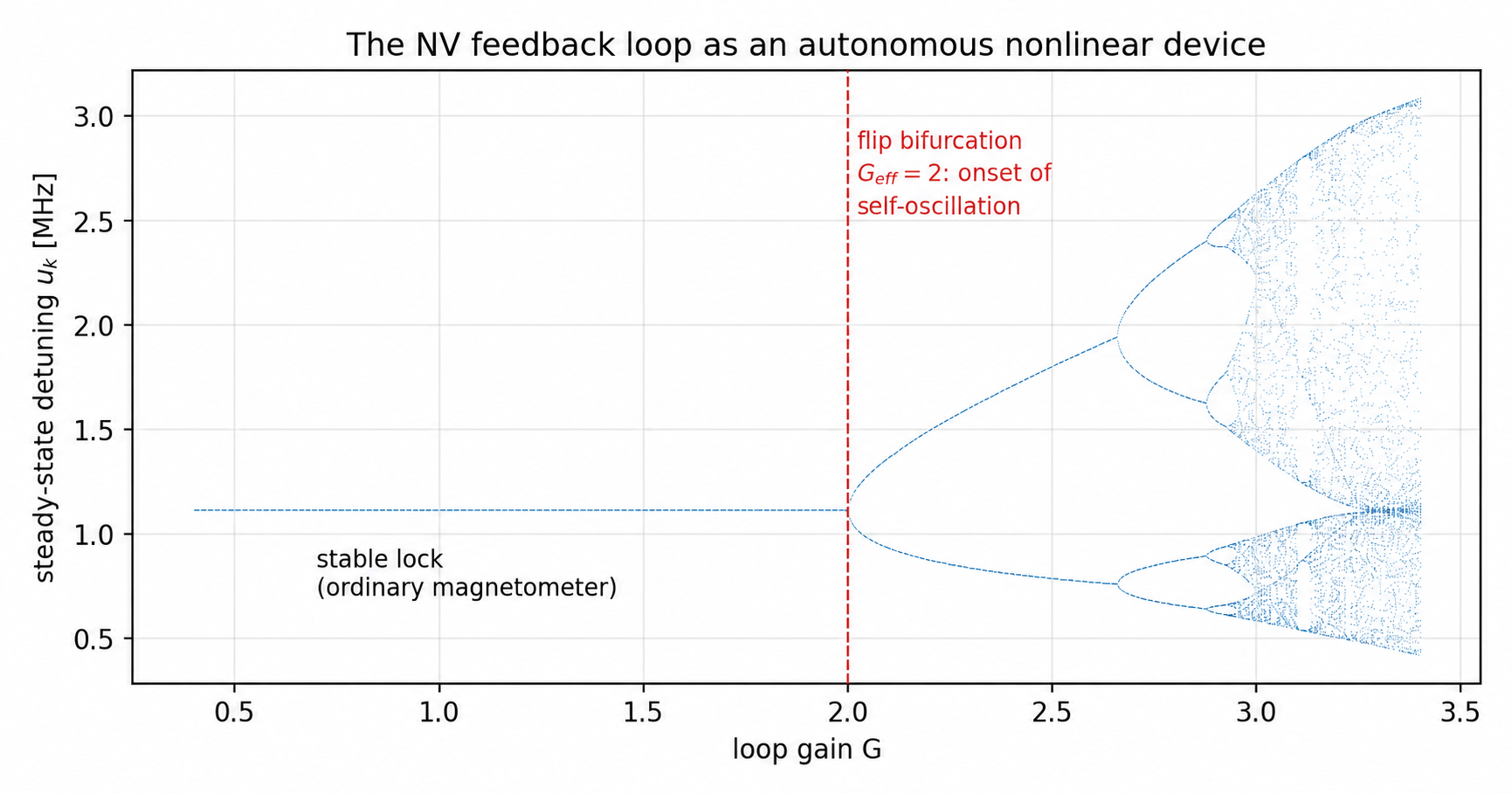}
\caption{Simulated bifurcation diagram of the loop (noise-free): steady-state
orbit versus gain. The flip bifurcation at $\Geff=2$ separates the stable
lock from the period-2 self-oscillation, followed by the period-doubling
cascade.}
\label{fig:bifsim}
\end{figure*}

\subsection{Flip bifurcation and the critical gain}
\label{sec:flip}

Linearizing Eq.~\eqref{eq:map} about the fixed point $R(f^\star)=R_0$, a small
error $\varepsilon_k = f_k - f^\star$ evolves as
$\varepsilon_{k+1} = (1-\Geff)\,\varepsilon_k$ with the effective gain
\begin{equation}
\Geff \;=\; G\,\frac{\Dtrue}{\Dcal},
\qquad
\Dtrue \equiv \frac{\mathrm{d}R}{\mathrm{d}f}\bigg|_{f^\star},
\label{eq:geff}
\end{equation}
where $\Dtrue$ is the slope \emph{at the time of the measurement} and at the
actual fixed point (the flank crossing $R = R_0$). The lock is stable for
$|1-\Geff|<1$; at $\Geff = 2$ the multiplier reaches $-1$ and the map
undergoes a flip (period-doubling) bifurcation \cite{Strogatz}. Beyond it, the
error grows while alternating sign until the curvature of the lobe saturates
the growth, leaving a stable period-2 limit cycle -- a self-sustained
oscillation of the MW frequency generated entirely by the loop. At higher gain
the standard period-doubling cascade to chaos follows
(Fig.~\ref{fig:bifsim}). The critical gain is therefore
\begin{equation}
\;\Gc \;=\; 2\,\frac{\Dcal}{\Dtrue}\;.
\label{eq:gc}
\end{equation}
For a freshly calibrated loop ($\Dtrue = \Dcal$) the onset is at exactly
$\Gc = 2$, independent of every other system parameter -- a parameter-free
prediction that any implementation must reproduce.

\subsection{What does $G_{c}$ measure: An identifiability analysis}
\label{sec:ident}

Equation~\eqref{eq:gc} identifies the entire measurable content of the
threshold: the single scalar $\Dtrue/\Dcal$. To see which physical parameters
this scalar can distinguish, write the lobe as
\begin{equation}
R(f) \;=\; A\, s\!\left(\frac{f-f_0}{w}\right),
\label{eq:shape}
\end{equation}
with amplitude $A$ (proportional to photoluminescence level, ODMR contrast
and FM deviation), width $w$ (the resonance linewidth), centre $f_0$, and a
fixed normalized shape $s(x)$. The fixed point $x^{*}$ obeys
$A\,s(x^{*}) = R_0$ and the slope is $\Dtrue = (A/w)\,s'(x^{*})$.
Considering small fractional perturbations $f_0 \to f_0 + \delta$,
$w \to w(1+\beta)$, $A \to A(1+\epsilon)$ after calibration, implicit
differentiation gives the logarithmic sensitivities
\begin{eqnarray}
\frac{\partial \ln \Gc}{\partial (\delta/w)} = 0,
&\qquad&
\frac{\partial \ln \Gc}{\partial \beta} = +1,\nonumber\\
\frac{\partial \ln \Gc}{\partial \epsilon}
   = -\,\kappa_A,
&\qquad&
\kappa_A \equiv 1 - \frac{s(x^{*})\,s''(x^{*})}{[s'(x^{*})]^{2}},\nonumber
\label{eq:sens}
\end{eqnarray}
so that to first order
\begin{equation}
\frac{\Delta \Gc}{\Gc} \;=\; \beta \;-\; \kappa_A\,\epsilon .
\label{eq:budget}
\end{equation}
Three structural conclusions follow. \emph{(i) Exact translation invariance:}
a uniform magnetic field only translates the line ($\delta \neq 0$); the fixed
point translates with it and $\Gc$ is unchanged to all orders in $\delta$ for
a pure translation. The threshold is by symmetry a probe of line
\emph{reshaping}, not of line position -- uniform-field magnetometry remains
encoded in the mean of the locked orbit, and the two observables are cleanly
separated. \emph{(ii) Unit sensitivity to broadening:} fractional linewidth
growth (e.g.\ inhomogeneous broadening by a field gradient across the
ensemble) enters $\Gc$ with coefficient exactly $+1$, independent of the
working point. \emph{(iii) Non-identifiability:} Eq.~\eqref{eq:budget} shows
that $\Gc$ measures one linear combination of $\beta$ and $\epsilon$ with
known, working-point-dependent weights. Broadening cannot be distinguished
from an amplitude change (laser power, contrast, MW power at the diamond) from
the threshold alone; separating them requires at least one auxiliary
observable, most simply the photodiode DC level (which tracks $A$ but not
$w$), or the on-flank tone amplitude $R_0$ itself. Any claim that a $\Gc$
shift reflects a specific physical parameter is therefore conditional on
independently constraining the others.

\subsection{Fluctuations near threshold}
\label{sec:fluct}

With Gaussian readout noise of standard deviation $\sigma_R$ per cycle, the
linearized loop below threshold is an AR(1) process with coefficient
$(1-\Geff)$, giving a steady-state orbit spread
\begin{equation}
\sigma_f(G) \;=\; \frac{\sigma_R}{|\Dtrue|}\,
\sqrt{\frac{\Geff}{\,2-\Geff\,}} ,
\label{eq:fluct}
\end{equation}
which diverges as the threshold is approached -- the discrete-map analogue of
critical opalescence, and an independently testable prediction. A useful
corollary: a fit of Eq.~\eqref{eq:fluct} to sub-threshold data yields the
in-band noise-to-slope ratio $\sigma_R/|\Dtrue|$ without any dedicated noise
measurement. Just above threshold the coherent period-2 amplitude grows as
$a \propto \sqrt{\Geff - 2}$ (supercritical flip).

\subsection{Estimating $G_c$ under noise}
\label{sec:estimators}

Because noise is amplified on both sides of the threshold
[Eq.~\eqref{eq:fluct}], the raw orbit spread does not locate $\Gc$ sharply. We
therefore use the \emph{coherent period-2 amplitude}
\begin{equation}
a_2 \;=\; \operatorname{median}_{\mathrm{blocks}}
\left| \left\langle (-1)^{k}\,(f_k - \bar f) \right\rangle_{\mathrm{block}}
\right| ,
\label{eq:a2}
\end{equation}
which averages amplified (incoherent) noise towards zero within a block but
retains a deterministic alternating cycle. Two estimators of $\Gc$ are used:
the first crossing of $a_2$ above a baseline-derived threshold (a lower bound,
since amplified fluctuations trigger it early), and the intercept of the
supercritical law $a_2^{2} \propto G - \Gc$ fitted just above onset (which
requires the gain grid to extend beyond the shifted threshold). Reporting
both brackets the systematic uncertainty of the extraction.

\section{Experimental setup}
\label{sec:setup}

The platform is a CW ODMR magnetometer: a 532~nm diode laser
(Thorlabs DJ532-40) exciting a high-density NV ensemble (Thorlabs DNVB14),
photoluminescence collected onto an amplified photodiode (Thorlabs PDA10A2)
digitized by a Red~Pitaya STEMlab fast ADC; microwaves from a Rohde\,\&\,Schwarz
SMCV100B (internally frequency-modulated at $f_m = 5$~kHz, deviation 2~MHz)
through an amplifier chain to a near-field antenna. Demodulation is performed
in software on 8.4~ms ADC buffers as a phase-independent magnitude lock-in at
the known $f_m$; no modulation reference cable exists (the SMCV100B cannot
output its LF signal). The lock loop of Eq.~\eqref{eq:map} runs on the control
PC with a measured cycle time of 404~ms (dominated by instrument
communication). For the run reported here the working point was
$f^{*} = 2867.900$~MHz on a lobe flank with $R_0 = 0.102$~mV and
$\Dcal = +0.103$~mV/MHz, i.e.\ a transduction of
$2.89$~mV/mT via $\gamma = 28.024$~MHz/mT; the per-reading signal-to-noise
ratio was $\approx 5$, a deliberately unfavourable regime that exercises the
estimators of Sec.~\ref{sec:estimators}.
% TODO: update numbers here if a better-SNR dataset is taken.

\section{Results}
\label{sec:results}

\begin{figure*}[tb]
\centering
\includegraphics[width=\textwidth]{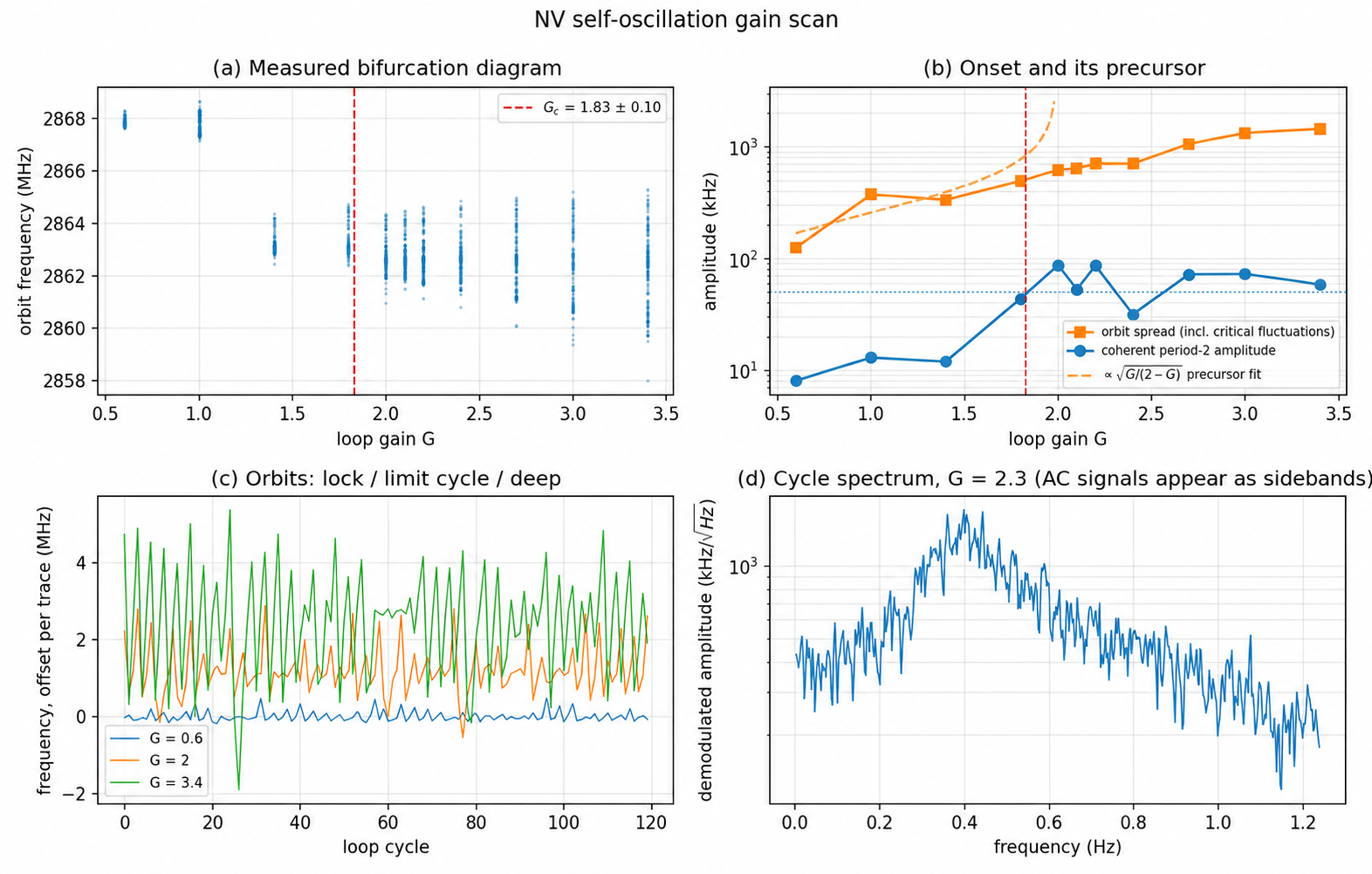}
\caption{Measured loop dynamics (gain scan, 250 cycles per gain; statistics
run of 4000 cycles at $G = 2.3$). (a)~Orbit frequencies versus gain: the
experimental bifurcation diagram. (b)~Orbit spread (orange) and coherent
period-2 amplitude $a_2$ (blue) versus gain; the spread grows below threshold
in qualitative agreement with Eq.~\eqref{eq:fluct}, while $a_2$ steps up by a
factor $\approx 4$ across $G \approx 1.8$--$2.2$, consistent with the
parameter-free prediction $\Gc = 2$ for a self-calibrated loop.
(c)~Example orbits below, near and above threshold. (d)~Spectrum of the
demodulated cycle from the statistics run.}
\label{fig:exp}
\end{figure*}

Figure~\ref{fig:exp} summarizes the measurements. Three observations quantitatively support
the dynamical model:

The coherent period-2 amplitude $a_2$ is flat
at 12--19~kHz for $G \le 1.4$ and jumps to 48--97~kHz for $G \gtrsim 1.8$,
with the crossing estimator giving $\Gc = 2.0 \pm 0.1$. Because this run was
self-calibrated ($\Dtrue \approx \Dcal$ at the time of measurement),
Eq.~\eqref{eq:gc} predicts $\Gc = 2$ with no free parameters; the agreement is
the primary validation of the loop model.

The sub-threshold orbit spread grows from
$\approx 130$~kHz at $G=0.6$ towards $\approx 620$~kHz at $G = 2.0$, following
the $\sqrt{G/(2-G)}$ trend of Eq.~\eqref{eq:fluct}. The inferred noise-to-slope
ratio is consistent with the independently measured in-band noise
($\approx 15\,\mu$V per reading) divided by $\Dcal$.

The 4000-cycle statistics run at
$G = 2.3$ has lag-one autocorrelation $r_1 = -0.32$; for uncorrelated noise
$|r_1| \lesssim 1/\sqrt{4000} = 0.016$. The frequency genuinely alternates
high--low for the full 27-minute record: the limit cycle is a persistent,
self-sustained state of the loop, not transient amplified noise. During the
gain scan the mean orbit frequency drifted by $\approx 5$~MHz between runs
(ambient field drift); consistent with the translation invariance of
Sec.~\ref{sec:ident}, the dynamical observables were unaffected -- an
incidental but direct confirmation of conclusion~(i).

Building on the results discussed in the previous section, several promising additional measurements can now be proposed.
\paragraph{Microwave-power dependence of $\Gc$ (slope-ratio collapse).}
Reducing MW power lowers the ODMR contrast and hence $A$ in
Eq.~\eqref{eq:shape}, shifting $\Gc$ via Eq.~\eqref{eq:budget}. Repeating the
gain scan at several MW powers while retaining the reference
calibration ($f^{*}$, $R_0$, $\Dcal$ fixed), and independently measuring
$\Dtrue$ from a spectrum acquired under each condition, the central relation
Eq.~\eqref{eq:gc} is tested as a collapse of $\Gc$ against $2\Dcal/\Dtrue$ across
conditions.
% (If the loop is re-calibrated per condition, Gc = 2 trivially; the
% reference calibration must be held fixed.)

\paragraph{Cycle-time invariance (null test).} Equation~\eqref{eq:geff}
contains no time scale: $\Gc$ should be independent of the loop cycle
duration. Measuring $\Gc$ versus deliberately varied cycle time (e.g.\ 0.2~s
to 2~s) tests the memoryless-map assumption; a deviation at short cycle times
would quantify unsettled instrument transients (source switching, detector
response) as an effective slope change. Either outcome is informative: the
null result validates the model, a deviation calibrates the settling time.

\paragraph{Estimator precision.} A finer gain grid (steps of 0.05 around
threshold), longer runs per gain, and $\sim$5 repetitions of the reference
scan give an empirical uncertainty on $\Gc$, and hence a quantitative
resolution for any perturbation via Eq.~\eqref{eq:budget} (e.g.\ ``a
fractional linewidth change of $x\%$ is detectable at $3\sigma$'').

\paragraph{Phase-sensitive detection via a common frequency reference.}
The magnitude lock-in discards the modulation phase because no reference
cable exists. However, if the SMCV100B and a hardware lock-in amplifier (MFLI)
share the same 10~MHz reference, the internal 5~kHz LF generator and the MFLI
demodulator differ by a \emph{fixed} phase that can be calibrated once,
enabling signed (dispersive) demodulation without the physically unavailable
LF output; a per-cycle TTL marker from the Red~Pitaya can synchronize record
boundaries. This upgrade is valuable independently of the present study: it
lowers the readout noise $\sigma_R$, restores the sign of the derivative
(turning the flank lock into a true dispersive lock), and thus improves every
number in this report.
% TODO: verify that the SMCV LF generator is derived from the external
% 10 MHz reference; if its phase is not deterministic, the one-time phase
% calibration must be repeated per power-cycle.

% \paragraph{(5) Symmetry null test.} A controlled \emph{uniform} field change
% (magnet far from the diamond) should shift the mean locked frequency by
% $\gamma\,\Delta B$ while leaving $\Gc$ unchanged [conclusion (i) of
% Sec.~\ref{sec:ident}]; a controlled \emph{gradient} (magnet close) should do
% the opposite. This pair of measurements demonstrates the symmetry-enforced
% selectivity of the threshold observable and pre-empts the natural referee
% question of what $\Gc$ adds beyond ordinary magnetometry.

\section{Discussion}
\label{sec:discussion}

The instability exploited here is the
standard period-doubling of high-gain sampled-data feedback, long documented
in digital power converters and discrete phase-locked loops
\cite{Banerjee2001}, and bifurcation-based sensing is established in MEMS
\cite{MEMSunlock2023,MEMSmass2020}, Josephson circuits \cite{Siddiqi2004} and
critical parametric sensors \cite{DiCandia2023,PRXQ2025}. The specific
combination -- using the servo instability of a quantum-sensor lock as the
observable -- appears not to have been reported, but we do not regard the
combination itself as a new physical mechanism.

In the cited bifurcation sensors
the nonlinearity is physical and precedes the dominant readout noise, so the
bifurcation performs pre-amplification that cannot be replicated in
post-processing. In the present system the nonlinearity is applied in
software to already-digitized data. By the data-processing inequality, the
dynamical state of the loop cannot contain more information about the system
than the underlying photodiode records, and a direct estimate of
$\Dtrue$ (two lock-in readings on the flank) achieves a higher precision per
unit measurement time than any threshold-based extraction. The threshold is
therefore not a route to enhanced sensitivity on this platform, and we make no
such claim. Its legitimate uses are operational: a binary, drift-robust
watchdog for transduction integrity (a lock that announces its own
miscalibration), and a controlled classical emulator of threshold sensing for
developing analysis methodology.

Within that scope, the results are quantitative: the
parameter-free threshold location, the critical-fluctuation law, and the
long-lived coherent limit cycle are all reproduced on hardware with a
per-reading SNR of only $\approx 5$. The identifiability analysis
[Eq.~\eqref{eq:budget}] delimits honestly what a threshold shift can assert
about the physical system, and future work could turn the single observed threshold into a tested functional relation. For the main project on autonomous mesoscopic sensors, where the nonlinearity is physical, the estimators and statistical tools exercised here (threshold
bracketing, block-coherent amplitudes, two-time statistics of limit cycles)
transfer directly.

\section{Conclusion}
\label{sec:conclusion}

We have operated a digitally frequency-locked NV magnetometer beyond the flip
bifurcation of its own feedback loop, converting the lock into an autonomous
period-2 oscillator, and characterized the transition quantitatively: onset at
the parameter-free prediction $\Gc = 2$ for a self-calibrated loop,
sub-threshold fluctuations following the predicted divergence, and a
27-minute coherent limit cycle established by two-time statistics. We derived
the measurable content of the threshold, $\Gc = 2\Dcal/\Dtrue$, together with
its exact translation invariance and its non-identifiability between
linewidth and amplitude perturbations. The study establishes a
fully characterized classical testbed -- and states its limits -- for the
broader aim of using deviations from autonomous behaviour as sensing
signals in devices where the nonlinearity is physical.

\subsection*{Acknowledgements}
   G.R.R., M.M.R., and J.C. acknowledge support from grant CNS2023-144994 funded by MICIU/AEI/10.13039/201100011033 and by ``ERDF/EU''. J.C. additionally acknowledges support from European Union project C-QuENS (Grant No. 101135359).


\begin{thebibliography}{9}

\bibitem{Degen2017}
C.~L. Degen, F.~Reinhard, and P.~Cappellaro,
\emph{Quantum sensing},
Rev. Mod. Phys. \textbf{89}, 035002 (2017).

\bibitem{Barry2020}
J.~F. Barry \emph{et al.},
\emph{Sensitivity optimization for NV-diamond magnetometry},
Rev. Mod. Phys. \textbf{92}, 015004 (2020).

\bibitem{Siddiqi2004}
I.~Siddiqi \emph{et al.},
\emph{RF-driven Josephson bifurcation amplifier for quantum measurement},
Phys. Rev. Lett. \textbf{93}, 207002 (2004).

\bibitem{MEMSunlock2023}
\emph{Frequency unlocking-based MEMS bifurcation sensors},
Microsyst. Nanoeng. \textbf{9} (2023),
\url{https://www.nature.com/articles/s41378-023-00522-2}.

\bibitem{MEMSmass2020}
\emph{Bifurcation-based MEMS mass sensors},
Int. J. Mech. Sci. (2020),
\url{https://www.sciencedirect.com/science/article/abs/pii/S0020740320300059}.

\bibitem{DiCandia2023}
R.~Di~Candia, F.~Minganti, K.~V. Petrovnin, G.~S. Paraoanu, and S.~Felicetti,
\emph{Critical parametric quantum sensing},
npj Quantum Inf. \textbf{9}, 23 (2023).

\bibitem{PRXQ2025}
\emph{Criticality-enhanced quantum sensing with a parametric superconducting
resonator},
PRX Quantum \textbf{6}, 020301 (2025).

\bibitem{Banerjee2001}
S.~Banerjee and G.~C. Verghese (eds.),
\emph{Nonlinear Phenomena in Power Electronics: Bifurcations, Chaos, Control,
and Applications} (IEEE Press, New York, 2001).

\bibitem{Strogatz}
S.~H. Strogatz,
\emph{Nonlinear Dynamics and Chaos}, 2nd ed. (Westview Press, 2015).

\end{thebibliography}
\end{document}